\documentclass[aps,prb,preprint,superscriptaddress,showpacs]{revtex4}
\begin{document}

\preprint{MRI-P-011004}

\title{Ultrasonic Attenuation in Clean d-Wave Superconductors}

\author{Tribikram Gupta }
\affiliation{Harish-Chandra Research Institute, Chhatnag Road,
 Jhusi, Allahabad 211019, India}
                   
  \author{D. M. Gaitonde}
  \affiliation{High Pressure Physics Division, 
  Bhabha Atomic Research Centre, Mumbai 400085, India}

\date{\today}

\begin{abstract}
We calculate the low temperature longitudinal ultrasonic attenuation rate $\alpha_S$
 in clean d-wave superconductors.  We consider the contribution of previously ignored processes 
involving the excitation of a pair of quasi-holes or quasi-particles. These processes, which are 
forbidden by energy conservation in conventional s-wave superconductors, have a finite phase
space in d-wave superconductors due to the presence of nodes in the gap which give rise to soft 
low-energy electronic excitations. We find the contribution to $\alpha_S$ 
from these processes to be proportional to $T$ in the regime $k_B T\ll Qv_{\Delta} \ll \Delta_0$,(ultra-low 
temperature regime) and to be  proportional to $1/T$ in the region $Qv_F \ll k_BT \ll \Delta_0$, (low 
temperature regime) where ${\bf Q}$ is the ultrasound wave-vector and $\Delta_0$ is the maximum gap amplitude. We 
explicitly evaluate these terms, 
for parameters appropriate to the cuprates, for ${\bf Q}$ along the nodal and the antinodal directions and
compare it with the contribution from processes considered earlier(I.Vekhter et al 
{\it Phys. Rev.}{\bf B59}, 7123(1999)). In the ultra-low temperature regime, the processes considered 
by us make a contribution which is smaller by about a factor of 10 for ${\bf Q}$ along the nodal direction, 
while along the antinodal direction it is larger by a factor of 100 or so. In the low temperature regime on 
the other hand the contribution made by these terms is small. However taken together with the original terms 
we describe a possible way to evaluate the parameter $v_F/v_\Delta$. 
\end{abstract}

\pacs{74.20.Rp; 74.25.Ld}

\maketitle

 Ultrasonic attenuation  has been an important probe
in the understanding of BCS superconductors.
One of the earliest verifications of the temperature dependence
of the gap \cite{Morse}, predicted by BCS theory, was made
by measuring the ultrasonic attenuation co-efficient (UAC). 
 Thus, it is   expected to shed light on the nature of the
superconducting state in the cuprates as well. The calculation of the low temperature
attenuation rate
in d-wave superconductors has been considered earlier, both within the
clean limit \cite{Vekh,Carb,MPLB} and in the opposite hydrodynamic limit
\cite{Coleman,new}. 
Even earlier ultrasonic attenuation in anisotropic superconductors
has received attention both in the clean \cite{Coppersmith}
and in the dirty \cite{hf1,hf2,hf3,hf4} limits.
On the experimental front, there exists just the one
experiment \cite{Shobo} but the results of our work, which have been obtained
in the limit $Ql\gg 1$, that is the clean limit (where $Q$ is the ultrasound wave-vector
and $l$ is the electronic mean free path) are inapplicable there.
Assuming the ultrasound frequency to be 150 MHz, which lies in the
range in which typical ultrasound experiments \cite{Bishop,Batlogg}
are done, and the sound velocity \cite{Shobo} to be $4\times 10^5 m/s$,
 the clean limit \cite{Vekh} corresponds to the condition $l\gg 4\mu m$. 
Thus our calculations are applicable only to the cleanest samples\cite{Hosseini}.
High purity YBCO crystals have been fabricated in $BaZrO_3$ crucibles \cite{erb}having $l\sim 4\mu m$,
giving rise to hope that the clean limit may be experimentally realizable.

There exists evidence \cite {Randeria} that although
the normal state in the cuprates remains a puzzle in the absence
of well-defined quasi particles, the superconducting state is well-described by
BCS-like quasi particles suitably adapted to the symmetry of
the order parameter in the system. Further, experimental evidence largely points towards a d-wave order 
parameter \cite{Leggett} in the high $T_c$ materials. The layered nature of the cuprate superconductors together with
strong electronic correlation leads to a highly anisotropic electronic transport in the normal state,
even though the superconducting state has three-dimensional phase coherence due to inter planar Josephson
coupling. However most superconducting properties can be described by assuming 2-d BCS quasi-particles that have
an energy gap with a $d_{x^2 - y^2}$ symmetry. For the purposes of this paper we assume that the d-wave 
superconducting quasi-particle energy dispersion to be phenomenological.
 
The quasi particles have a gap function
that has nodes on the Fermi surface at the points $k_x=\pm k_y$.  
At low energies, the relevant low energy electronic excitations, are the Dirac quasiparicles
near these nodes. These nodal quasi-particles have energies that are well described
by $E^0_{\bf k}=\sqrt{v_F^2k_1^2+v_{\Delta}^2k_2^2}$. 
This Dirac spectrum is easily obtained \cite {Lee} by a linear expansion
of the band energy and gap about the nodes of the single particle excitation spectrum.
Here $k_1$ and $k_2$ are the momentum components normal and tangential to the Fermi surface at 
the nodal points. 
While it is possible to obtain such a quasiparticle spectrum from 
a suitably chosen microscopic Hamiltonian \cite{Arun,MPLB} in this
work we treat $v_F$ and $v_{\Delta}$ as phenomenological parameters whose values
should be read from experiments. Because of the dominance of the band energy scale 
$W > {\Delta_0}$ , where W is the band width and $\Delta_0$ is the maximum amplitude of the
superconducting gap, $v_F\over v_\Delta$ is large 
and is approximately 10 for parameters relevant to the cuprates. 
  
The UAC is the inverse of the length obtained by a product of the sound velocity
and the phonon life-time and sets the scale over which an ultrasonic wave decays while propagating
in the solid. The temperature dependent variation of the sound velocity
is very weak in the temperature range of interest to us. We therefore
calculate only the dependence of the phonon life-time on temperature. In general the 
damping of sound waves in the lattice can arise due to various excitations
in the system. Here we focus exclusively on the damping of phonons due to their 
interaction with  electrons. We ignore the 
contribution of anharmonic terms in the phonon Hamiltonian, as these are unaffected by the 
onset of superconductivity. Thus, assuming the usual electron-phonon coupling, the  UAC is 
proportional to the imaginary part of the electronic density-density correlation function.

A  calculation of the low temperature UAC in d-wave superconductors was first 
performed   by Vekhter et. al. \cite{Vekh}
and generalized to the orthorhombic situation by Wu and Carbotte \cite{Carb}.
They obtained an attenuation rate which is proportional to the temperature
at $T\ll \Delta_0$. To obtain a result that is independent of the poorly known
electron-phonon coupling constant, they normalise their result by dividing
the low temperature attenuation rate ($\alpha_S(T)$) by the attenuation rate 
at $T_c$. In order to obtain a result analytically for $\alpha_S(T_c)$,
it is necessary to assume a parabolic band dispersion. It is by now well-established
\cite{Norman} that the Fermi surface in the cuprates deviates strongly from circular
symmetry. Thus there is an unknown , Fermi surface sensitive wave vector dependence in 
$\alpha_S(T_c)$, which has been missed.
A fundamental drawback of their calculation is that the use of a quasiparticle
description at $T_c$ is itself questionable given the anomalous nature of the
normal state in the cuprates.

Apart from these questions, all previous studies of the UAC in references Refs.(2,3,4,7) ignore
the contribution of processes corresponding to the creation of
a pair of quasi particles or a pair of quasi holes.
In conventional s-wave superconductors, where the condition $\Omega\ll \Delta$
is satisfied, energy conservation forbids these processes. However, in d-wave superconductors
the nodes in the energy gap imply that there is a finite phase space for these processes. In Ref.(3) it is 
claimed that phase space restrictions make the contributions accruing to the UAC from here
negligibly small. In this paper we explicitly evaluate the contributions
of these processes in the low temperature regime. 
At ultra low temperatures ($k_BT\ll Qv_{\Delta}\ll\Delta_0$) where $Q$
is the ultrasound wave-vector, $v_F$ is the Fermi velocity at the nodes
and $\Delta_0$ is the maximum amplitude of the superconducting gap, 
we find that there is a term in $\alpha_S$ linear in $T$ from these processes
in addition to a similar contribution evaluated earlier by Vekhter et. al.
However, the co-efficient of the linear in $T$ term, in this case
is found to be, for parameters appropriate to the cuprates, larger
by a factor of 100 for sound wave along the antinodal direction. In the nodal direction these new terms provide
corrections that are smaller by a factor of 10. 
    
The other regime we consider is $Qv_F\ll k_BT\ll \Delta_0$ and here we find  the contribution to the UAC from these
previously neglected processes is proportional to $1/T$. Using the 
ratios of the co-efficients of the $1/T$ contribution calculated
by us and the linear in $T$ contribution evaluated in Ref.(2) 
for two different ultrasound wave-vectors ${\bf Q}_1$ and ${\bf Q}_2$
it is possible to determine the ratio $v_F/v_{\Delta}$  which is a basic parameter in the description of the 
superconducting state in the cuprates.

The imaginary part of the density-density response function is given by
$\chi^"({\bf Q},\Omega)=\chi^"_a({\bf Q},\Omega)+\chi^"_b({\bf Q},\Omega)$.
Here 
$\chi^"_a({\bf Q},\Omega)=X1-X2$ and $\chi^"_b({\bf Q},\Omega)=X3-X4$, where
\begin {equation}
X1= {-2\pi\over N}\Sigma_{\bf k}[n(E_{\bf k})-n(E_{{\bf k}+{\bf Q}})]
(u_{{\bf k}+{\bf Q}}^2u_{\bf k}^2-{\Delta_{\bf k}\Delta_{{\bf k}+{\bf Q}}\over 4 E_{\bf k}
E_{{\bf k}+{\bf Q}}})\delta(\Omega+E_{\bf k}-E_{{\bf k}+{\bf Q}}),
\end{equation}
\begin {equation}
X2={-2\pi\over N}\Sigma_{\bf k}[n(E_{\bf k})-n(E_{{\bf k}+{\bf Q}})]
(v_{{\bf k}+{\bf Q}}^2v_{\bf k}^2-{\Delta_{\bf k}\Delta_{{\bf k}+{\bf Q}}\over 4 E_{\bf k}
E_{{\bf k}+{\bf Q}}})\delta(\Omega-E_{\bf k}+E_{{\bf k}+{\bf Q}})
\end{equation}
\begin{equation}
X3={-2\pi\over N}\Sigma_{\bf k}[1-n(E_{\bf k})-n(E_{{\bf k}+{\bf Q}})]
(u_{{\bf k}+{\bf Q}}^2v_{\bf k}^2+{\Delta_{\bf k}\Delta_{{\bf k}+{\bf Q}}\over 4 E_{\bf k}
E_{{\bf k}+{\bf Q}}})\delta(\Omega-E_{\bf k}-E_{{\bf k}+{\bf Q}}),
\end{equation}
and 
\begin{equation}
X4={-2\pi\over N}\Sigma_{\bf k}[1-n(E_{\bf k})-n(E_{{\bf k}+{\bf Q}})]
(v_{{\bf k}+{\bf Q}}^2u_{\bf k}^2+{\Delta_{\bf k}\Delta_{{\bf k}+{\bf Q}}\over 4 E_{\bf k}
E_{{\bf k}+{\bf Q}}})\delta(\Omega+E_{\bf k}+E_{{\bf k}+{\bf Q}}).
\end{equation}

To proceed further, we retain terms only to the leading order in ${\bf Q}$
and $\Omega$. To this order $\chi^"_a$ reduces to
\begin{equation}
\chi^"_a= {-2\pi\Omega\over N}\Sigma_{\bf k}{n'(E_{\bf k})}\delta(\vec{Q}.\vec{v_g}){\xi_{\bf k}^2\over E_{\bf k}^2}
\end{equation}
where $\vec{v_g} = \nabla_{\bf k}E_{\bf k}$.
  
This term has been calculated earlier by the authors of Ref.(2).
For $k_BT\ll \Delta_0$ , it yields
\begin{equation}
\chi^"_a={-4ln2\Omega a^2\over 2\pi}{k_BT\over v_Fv_{\Delta}}
[{\beta^2\over (\alpha^2+\beta^2)^{3/2}}+{\lambda^2\over (\lambda^2+\eta^2)^{3/2}}]
\end{equation}
Here $\alpha=v_FQ\cos(\pi/4-\theta)$, $\beta=v_{\Delta}Q\sin(\pi/4-\theta)$,
$\eta=v_FQ\sin(\pi/4-\theta)$ and $\lambda=v_{\Delta}Q\cos(\pi/4-\theta)$.
We now turn our attention to the remaining contribution to $\chi"$
which comes from the terms in Eqs. (3) and (4). In conventional s-wave superconductors
we have $\Omega\ll 2\Delta$ and so the energy conserving delta function
in Eqs. (3) and (4) can never be satisfied. Thus $\chi_b^"({\bf Q},\Omega) = 0$ 
in that case. However, as mentioned earlier, the presence of nodes in the gap
function leads to a finite contribution from $\chi_b^"({\bf Q},\Omega)$ in 
d-wave superconductors.  It has always been assumed in previous work
\cite{Coppersmith,JKB,Swihart,Vekh,Carb,MPLB} that the phase space available for the processes described by this term (creation or annihiliation of a pair of quasi particles)
would be small and so its contribution has been ignored. In this
paper we explicitly evaluate this term. We find that its contribution is large
and cannot be ignored at ultra low($T\ll 1mK$) temperatures.  
In the low temperature regime ($1mK\ll T \ll \Delta_0/k_B$) although the relative size of the terms evaluated by us
is small its different ($1/T$) temperature dependence enables the determination of the parameter $v_F/v_{\Delta}$ 
by measuring the attenuation rate at 2 different angles.
These are the principal results of this paper.

We now present our calculation of $\chi_b^"({\bf Q},\Omega)$. We work in units of lattice constant $a=1$.
To linear order in $\Omega$, we find
\begin{equation}
\chi_b^"({\bf Q},\Omega) \cong {2\pi\Omega\over N}\Sigma_{\bf k}{n'(E_{\bf k})}({u_{{\bf k} + {\bf Q}}}^2{v_{\bf k}}^2 + {v_{{\bf k} + {\bf Q}}}^2{u_{\bf k}}^2 + {\Delta_{\bf k}\Delta_{{\bf k}+{\bf Q}}\over 2 E_{\bf k}
E_{{\bf k}+{\bf Q}}})\delta(E_{\bf k}+E_{{\bf k}+{\bf Q}})
\end{equation}
To leading order in Q, $\chi_b"$ further reduces to 
\begin{equation}
\chi^"_b({\bf Q},\Omega) \cong {2\pi\over N}\Sigma_{\bf k}{\Omega}n'({E_{\bf k}})\delta(2E_{\bf k}+ \vec{Q}.\vec{v_g})\
({{\Delta_k}\over{E_k}})^2
\end{equation}

Linearizing the band energies and the gap function at the nodes \cite{Lee} which lie at the intersection of the lines 
$k_x$ = $\pm k_y$ with the Fermi surface, we find that $\xi_{\bf k} \approx v_{F}k_1$, $\Delta_{\bf k}$ = $v_{\Delta}k_2$ and
$E_{\bf{k}}$ = $\sqrt{v_{F}^2k_1^2 + v_\Delta^2k_2^2}$ where $k_1$ and $k_2$ are the components of the momenta normal and tangential to the Fermi surface at the nodes. We take ${\bf Q}$ = $Qcos\theta{\hat x} + Qsin\theta{\hat y}$. Going over 
to polar coordinates ($v_{F}k_1$ = $\rho{cos{\phi}}$ and $v_{\Delta}k_2$ = $\rho{sin{\phi}}$) we find that for the
node near ($\pi/2,\pi/2$), 
${\bf{v_g}}.{\bf Q}$  = ${\alpha}cos{\phi}
+ {\beta}sin{\phi}$.

In comparison to the corresponding contribution for $\chi_a"$ an extra term $2\rho$ appears in the argument of the 
delta function in Eqs.(3) and (4). Finally we obtain for the node near $(\pi/2,\pi/2)$, 
\begin {equation}
\chi^"_{b1}({\bf Q},\Omega) = {\Omega\over {2{\pi}v_Fv_{\Delta}}} {{\int}_{0}^{\tilde \rho_c}}d\rho\rho{n'(\rho)}
\int_0^{2\pi}{d\phi}\delta(2\rho + {\alpha}cos{\phi} + {\beta}sin{\phi}){sin^2{\phi}}\\
\end {equation}
where $\tilde \rho_c$ = $\sqrt{\pi v_Fv_\Delta}$ is a cutoff chosen to preserve the volume of the Brillouin zone.
We perform the $\phi$ integral first. The condition that the zeros of $2\rho + {\alpha}cos{\phi} + {\beta}sin{\phi}$
are real together with the requirement that $cos^2\phi_0^{\pm} \leq 1$ (where $\phi_0^{\pm}$ are the 2 independent
zeroes in the range $0 \leq \phi \leq 2\pi$) leads to the constraint $\rho \leq {\sqrt{\alpha^2 + \beta^2}\over 2}$
= $\rho_c$. Assuming $Q\approx 400 cm^{-1}$ and ${v_\Delta/v_F} \approx {1\over k_F \xi} \approx 1/10$ for the 
cuprates  we find that $\rho_c \ll {\tilde \rho_c}$. Thus the contribution to $\chi_b"$ from this node can be 
written as     
 
\begin{equation}
\chi^"_{b1}({\bf Q},\Omega) = {\Omega\over {2{\pi}v_Fv_\Delta}}\int_0^{\rho_c}d\rho\rho{n'(\rho)}
{{{{8\rho^2(\beta^2 - \alpha^2)}\over {(\alpha^2 + \beta^2)^2}} + {2\alpha^2\over {\alpha^2 + \beta^2}}}
\over {(\alpha^2 + \beta^2 - 4\rho^2)^{1/2}}}
\end{equation}  

The change of the upper cutoff from ${\tilde \rho_c}$ to $\rho_c$ is the phase space reduction referred to 
by the authors of Ref.3.

We evaluate the expression in Eq.(10) in two regimes. a)$k_BT\ll Qv_{\Delta}\ll \Delta_0$
and b)$Qv_F\ll k_BT\ll \Delta_0$. The maximum value of $\rho_c$ is $\rho_c$(max) = 
$v_{F}Q/2$ and minimum value of $\rho_c$ is $\rho_c$(min) = $v_{\Delta}Q/2$. We have
taken the temperature regimes to be either a) much less than $\rho_c$(min) or b) much  greater 
than $\rho_c$(max). 

We approximate the derivative of the Fermi distribution
so as to simplify the integration and render it doable analytically. We notice that $n'(\rho)
\sim -1/4k_BT$ for $\rho/2k_BT \ll 1$ and $n'(\rho)$ $\sim {-e^{-\rho/k_BT}\over k_BT}$
for $\rho/2k_BT \gg 1$. We therefore make the interpolating approximation $n'(\rho) \approx 
-1/4k_BT$ for $\rho < 2k_BT$ and $n'(\rho) \approx {-e^{-\rho/k_BT}\over k_BT}$ for 
$\rho > 2k_BT$. Our approximation is asymptotically exact and is therefore expected to
provide reasonable answers in the 2 limits ($k_BT \ll \rho_c \ll \Delta_0$ and $\rho_c \ll k_BT \ll
\Delta_0$) considered by us.  

We first consider the ultra low temperature regime $k_BT \ll Qv_\Delta$. For typical parameters 
appropriate to the cuprates and assuming $Q\approx 400 cm^-1$ this condition implies 
$k_BT \ll 1 mK$. In this case $\chi_{b1}''$ can be split into 2 pieces and can be written as
\begin{equation}
\chi_{b1}''({\bf Q},\Omega) \cong (-1/4k_BT){\Omega\over \pi v_Fv_\Delta}[I_1 + I_2]
\end{equation}  
where
\begin{equation}
I_1 = {\int_0^{2k_BT}d\rho}{\rho\over {\sqrt{\alpha^2 + \beta^2 - 4\rho^2}}}
({4\rho^2(\beta^2 - \alpha^2)\over {(\alpha^2 + \beta^2)^2}} + {\alpha^2\over {\alpha^2 + \beta^2}})
\end{equation} 
and 
\begin{equation}
I_2 = 4\int_{2k_BT}^{\rho_c}d\rho{\rho{e^{-\rho\over {k_BT}}}\over {\sqrt{\alpha^2 + \beta^2  - 4\rho^2}}}
({{4\rho^2(\beta^2 - \alpha^2)}\over {(\alpha^2 + \beta^2)^2}} + {\alpha^2\over {\alpha^2 + \beta^2}})
\end{equation}
The integral in $I_1$ is elementary. To evaluate $I_2$ we further make the approximation 
${\int_l^u}dxe^{-x}f(x) \approx  e^{-l}\int_l^{min(l+1,u)}dxf(x)$  (for $u > l > 0$)
and where f(x) is a slowly varying function compared to $e^{-x}$. We similarly evaluate the 
contributions of the other three nodes as well. The final answer is a long expression 
and in the interest of clarity and brevity we only present the leading term in an expansion
in the small parameter ${k_BT\over \sqrt{\alpha^2 + \beta^2}}$. To this leading order we find
\begin{equation}
\chi_b'' = {-k_BT\Omega\over \pi v_Fv_\Delta}(1 + 5e^{-2})[{\alpha^2\over (\alpha^2 + \beta^2)^{3/2}}
+{\eta^2\over (\eta^2 + \lambda^2)^{3/2}}]     
\end{equation}
It is instructive to compare this result with the processes considered earlier in Eq.(6).
We find that upto factors of O(1), the result is the same with $\alpha \leftrightarrow \beta$
and $\lambda\leftrightarrow \eta$. This change comes from the different factor ${\Delta_{\bf k}^2
\over E_{\bf k}^2}\sim sin^2\phi$ in Eq. (8) as compared to ${\xi_{\bf k}^2\over
 E_{\bf k}^2}\sim cos^2\phi$ in Eq. (5). Also notice that at ultralow temperatures the
dimensionless cutoff to the integral $\rho_c/k_BT\gg 1$ so that the phase-space reduction
due to the replacement of ${\tilde \rho_c}$ by $\rho_c$ is ineffective as most of the contribution to the integral comes from the region where $\rho/k_BT\ll \rho_c/k_BT $. To get an idea of the
relative size of the contribution of the processes considered by us with respect to those 
considered earlier by Vekhter et. al. \cite{Vekh} we have evaluated ${\chi_b''\over \chi_a''}$
in the ultra low temperature regime for ${\bf Q}$ directed along a node and an antinode respectively.
In the nodal case we have $\theta=\pi/4$, $\alpha=v_FQ$, $\beta=\eta=0$ and $\lambda=v_{\Delta}Q$.
Then from Eqs. (6) and (14) we find that
\begin{equation} 
{\chi_b''\over \chi_a''}={(1+5e^{-2})\over 2ln2}{v_{\Delta}\over v_F}
\end{equation}
Thus the contribution from these terms is smaller by a factor of $v_{\Delta}/v_F\sim 1/10$.
For the antinodal direction we have 
$\theta=0$, $\alpha=\eta=v_FQ/{\sqrt 2}$, $\beta=\lambda=v_{\Delta}Q/{\sqrt 2}$.
Putting these values in Eqs. (6) and (14) we have
\begin{equation} 
{\chi_b''\over \chi_a''}={(1+5e^{-2})\over 2ln2}{v_{F}^2\over v_{\Delta}^2}
\end{equation}
Thus the ratio in this case is approximately 100 and so in this case it is the processes
calculated by us that make the dominant contribution.

We now proceed to the evaluation of $\chi_b''$ in the limit $Qv_F \ll k_BT \ll \Delta_0$.
In this case, we have $n'(\rho) \simeq -1/4k_BT$ over the entire range of $\rho$ integration
from 0 to $\tilde{\rho_c}$. Thus evaluating and summing the contribution to $\chi_b''$ from each
of the nodes we arrive at the result
\begin{equation}
\chi"_b = -{{\Omega}\over{{24\pi}v_{F}v_{\Delta}}}{1\over{k_{B}T}}
({{{\alpha}^2 + {2\beta}^2}\over ({\alpha}^2 + {\beta}^2)^{1/2}} + 
{{{\lambda}^2 + {2\eta}^2}\over ({\lambda}^2 + {\eta}^2)^{1/2}})
\end {equation} 

In this case the relative size of $\chi_b''$ with respect to $\chi_a''$ is very small
$O((Qv_F/k_BT)^2)$. However we now show how our result may be used to extract the parameter 
$v_F\over v_{\Delta}$. The total imaginary part of the density-density correlation function
can be written as 
\begin{equation}
\chi''_{total}({\bf Q},\Omega) = \chi_a''({\bf Q},\Omega) + \chi_b''({\bf Q},\Omega) = 
\gamma_1({\bf Q},\Omega)T + {\gamma_2({\bf Q},\Omega)\over T}
\end{equation}
where $\gamma_1$ and $\gamma_2$ are coefficients whose values are given in Eqs.(6) and (17).
Thus the UAC is given by
\begin{equation}
\alpha_S({\bf Q}, T) = M({\bf Q})[\gamma_1({\bf Q},\Omega)T + {\gamma_2({\bf Q},\Omega)\over T}]  
\end{equation}
where $ M({\bf Q})$ is a factor dependent on the electron-phonon coupling constant, the sound velocity 
and the ultrasound wave-vector. A measurement of $\alpha_S({\bf Q},T)$ and a comparision
of the two terms contributing to it would enable an extraction of the parameter 
$X({\bf Q}) = {\gamma_2({\bf Q}, \Omega)\over \gamma_1({\bf Q}, \Omega)}$. The poorly known
factor $M({\bf Q})$ exactly cancels out. On taking  the ratio of $X({\bf Q})$ evaluated at 
two different wave vectors we obtain an expression that depends only on the parameter
 $v_F\over v_{\Delta}$ and the wave vectors ${\bf Q_1}$ and ${\bf Q_2}$ respectively.
For the sake of illustration we have computed $X_{node}({\bf Q_1})/X_{antinode}({\bf Q_2})$
for ${\bf Q_1}$ along the nodal direction and ${\bf Q_2}$ along the antinodal direction. We find 
\begin{equation}
X_{node}({\bf Q_1})/X_{antinode}({\bf Q_2}) = {{2Q_1^2/Q_2^2}{(v_F/ v_\Delta)(1 + 2 v_\Delta/v_F)}\over 
{((v_F/v_\Delta)^2 + 1)((v_F/v_\Delta)^2 + 2)}}
\end{equation}
Thus the measurement of $\alpha_S({\bf Q},T)$ in the regime ($10^{-3} K\ll k_BT\ll \Delta_0$) provides an 
useful method to extract $v_F/v_\Delta$ without use of $\alpha_S({\bf Q}, T_c)$ whose
calculation necessitates assumptions about the existence of quasi-particles at $T_c$ with a 
circular Fermi surface.

Finally let us recapitulate the main points of this paper. We have calculated the UAC
for a d-wave superconductor. We have included the effects of processes involving
creation or annihilation of a pair of quasi-particles which are non zero because of the soft
 excitation spectrum of a d-wave superconductor. In the ultra-low temperature regime we 
find that these processes make a substantial contribution, which is linear in $T$, to the   
UAC. In the low temperature regime we find a small $1/T$ contribution to the UAC. We indicate 
how this can be used to extract the parameter $v_F/v_\Delta$. A drawback of our work is that
we have carried out calculations in the clean limit which makes it inapplicable to all but
the cleanest of samples. Yet another drawback is that we assume the existence of 
quasi-particles of spectral weight unity, thus neglecting all effects coming from 
the incoherent part of the spectral function seen in ARPES experiments \cite{Randeria}.

\end{document}